\documentstyle[onecolumn,psfig]{mn}
\newif\ifAMStwofonts
\newcommand       \beq          {\begin{equation}}
\newcommand       \eeq          {\end{equation}}

\newcommand       \cm           {\,{\rm cm}}

\newcommand	  \g		{\,{\rm g}}

\newcommand{\figwidth}{6.0in}

\title{Modelling the Astronomical Silicate Features:\\ 
I. On the Spectrum Subtraction Method}

\author[A. Li, J.M. Greenberg, \& G. Zhao]
       {Aigen Li$^{1,2,3,}$\footnotemark, 
        J. Mayo Greenberg$^{4}$, and Gang Zhao$^{3}$\\
         $^{1}$ Theoretical Astrophysics Program,
                University of Arizona, Tucson, AZ 85719, USA\\ 
         $^{2}$ Princeton University Observatory, Peyton Hall,
                Princeton, NJ 08544, USA\\
         $^{3}$ National Astronomical Observatories, Chinese Academy
                of Sciences, Beijing 100012, P.R. China\\
         $^{4}$ Raymond and Beverly Sackler Laboratory for Astrophysics,
                University of Leiden, Postbus 9504, 2300 RA Leiden, 
                The Netherlands}
\date{}

\pagerange{\pageref{firstpage}--\pageref{lastpage}}
\pubyear{}

\begin{document}

\maketitle

\label{firstpage}

\begin{abstract}
The {\it assumption of additive absorptivity} by 
different components in compound particles is a widely used
method applied in the literature to the analysis of the chemical
and structural properties of astronomical (circumstellar, 
interstellar, protostellar, and cometary) silicates as well
as other materials. The errors intrinsic in this additivity 
assumption, which, in application to astronomical spectra,
amounts in some case to {\it spectrum subtraction} have not always been 
adequately considered in previous works on silicate mineralogy.
The failings in the ``{\it spectrum subtraction method''} 
(intrinsically the same as the additive absorptivity assumption) 
are discussed here in terms of silicate core-ice mantle grains with 
various shapes. It is shown that these assumptions result in 
substantial errors for spherical grains. For spheroidal grains,
the errors are less significant and the spectrum subtraction 
method can be used to remove the ice mantle effects.
It is demonstrated that there is no significant improvement 
by considering a distribution of spheroidal shapes. 
It is further shown that the presence of additional 
organic mantles substantially modifies 
the silicate mineralogy interpretation. 
\end{abstract}

\begin{keywords}
ISM: dust, extinction -- infrared: ISM: lines and bands -- scattering
\end{keywords}

\section{Introduction}
\footnotetext{e-mail: agli@lpl.arizona.edu}

Silicate grains are ubiquitously present in various cosmic environments.
Theoretical efforts have been continuously made to identify the detailed 
chemical and structural properties of astronomical silicates (Draine \& Lee 
1984; Ossenkopf, Henning, \& Mathis 1992; Greenberg \& Li 1996; Mathis 1998; 
Demyk et al.\ 1999). On the other hand, during the last decade there has  
been considerable progress in laboratory measurements of astronomical 
silicate analogues (see Dorschner 1999 for a review). More recently, 
the high quality spectroscopic observations in the 10, 18$\mu$m 
(as well as the longer wavelength range) silicate bands region of 
a wide range astronomical objects by the {\it Infrared Space 
Observatory} (ISO) together with the rich laboratory data set 
offer us an unprecedented opportunity to study silicate dust mineralogy. 

The analysis of astronomical silicate composition is
commonly based on the direct comparison of astronomical 
absorption/emission features with laboratory data of terrestrial 
silicate candidates or with theoretical 
spectra calculated from the optical constants measured for silicate 
analogues. However, in most regions silicates are combined with 
other materials. For example, Greenberg (1978) proposed that in the diffuse
interstellar medium silicates are coated by a layer of carbonaceous organic
refractory; whereas Mathis \& Whiffen (1989) suggest that interstellar 
silicates are combined together with amorphous carbon grains to form 
porous composite aggregates.
GEMS (silicate glass with embedded metal and sulfides) in interplanetary
dust particles (IDP) which provide a spectral match to the interstellar
10$\mu$m Si-O stretching band is usually mantled with or embedded in
amorphous carbonaceous material (Bradley et al.\ 1999). 
Cometary dust particles are also believed to be a mixture of various forms
of silicate and organic (``CHON'' particles) materials (Greenberg 1982;
Kissel et al.\ 1986; Wooden et al.\ 1999).
In dense molecular clouds or protostellar disks, the consensus view is 
that refractory grains are further coated by an outer ice mantle.
 
In studying silicate composition, it is thus critical to accurately remove 
the effects caused by the grain's heterogeneous nature. A simple and 
commonly adopted method is to simply {\it subtract} the non-silicate 
contributions (determined from their characteristic absorption/emission 
properties) from the entire observational spectrum and then make a
comparison between the residual spectrum either directly with laboratory 
data or with a theoretically derived absorption spectrum by particles 
using the laboratory optical constants. 
Hereafter we term this the ``{\it spectrum subtraction}'' method.

For example, in a detailed effort to infer the precise composition of 
the silicate dust around two massive protostellar objects RAFGL 7009S 
and IRAS 19110+1045, Demyk et al.\ (1999) subtracted the ice absorption 
from the total observed absorption to obtain what they
presumed to be the observed silicate absorption spectrum as if it could 
be obtained by adding ``ice'' absorptivity to the absorption spectrum for 
bare spherical silicate material grains. Two key assumptions here, 
both of which affect the derived absorption, are that the grains are 
spherical and that the absorption by a core-mantle particle is the same
as the absorption by the core plus the absorption by the mantle. We term
this the ``{\it absorption additivity assumption}''. 
Note the spectrum subtraction method and the absorption additivity 
assumption are intrinsically the same. On the basis of this method, 
Demyk et al.\ (1999) then modelled the observed silicate absorption 
spectrum by performing Mie calculations (together with effective medium 
theory) of spherical porous grains of various compositions. 

In the current studies of the crystallinity of circumstellar silicate 
grains, the circumstellar dust emission spectra are often modelled as
co-added emissivities of individual dust components in spite of the 
fact that (at least) some dust components are actually not separated 
(e.g., Malfait et al.\ 1998). The absorption additivity approach has 
also been applied to cometary silicate analysis 
(see e.g. Brucato et al.\ 1999).

The absorption additivity (spectrum subtraction) method 
takes as a {\it premise} that the coupling between the silicate 
and non-silicate materials is negligible. 
It is obvious that such a premise is physically incorrect
(e.g. see Bohren \& Huffman 1983).
However, this fundamental question has not always been adequately 
considered in modelling the silicate emission/absorption 
features in the literature. 
An equally important problem has to do with the effect
of particle shape on the absorption spectrum.

The purpose of this work is not to model any specific 
astronomical objects, but to address the question 
of to what extent or in what circumstances such a premise 
is valid or acceptable. 
In doing so we will consider a simple case: silicate grains coated by 
an ice mantle as frequently proposed in the literature although it is 
likely that an inner carbonaceous mantle may also be present 
(Greenberg 1978). 
In \S2 we summarize the optical constants employed 
in this paper. In \S3 we discuss the spectrum subtraction method in terms 
of spherical shape. \S4 is devoted to spheroidal shape. 
We extend then to a distribution of shapes in \S5.
In \S6 we discuss the implications of a more realistic model; i.e., 
the coated core-mantle spheroidal grains.
We summarize our results in \S7.
In a subsequent paper we will perform detailed modelling on the interstellar
silicate bands and discuss the depletion of heavy elements and its implication
on dust models (Zhao, Li, \& Greenberg 2001).

\section{Optical properties of dust materials}
We consider three types of dust materials: amorphous silicates, carbonaceous
organic refractories, and ices. We take the dielectric functions of amorphous
olivine MgFeSiO$_4$ and organic refractory materials from Li \& Greenberg 
(1997). For ices, we consider a mixture of 
${\rm H_2O: CH_3OH: CO_2 = 1.3: 1: 1}$ (80-90K) which provides the best 
interpretation of the ISO spectra of RAFGL 7009S (Dartois et al.\ 1999).
The incorporation of CH$_3$OH and CO$_2$ adds substantial ice mantle
effects to the silicate features since they have strong absorptions such as
the 9.75$\mu$m C-O stretching mode of CH$_3$OH, and the 15.2$\mu$m C-O
bending mode of CO$_2$. We note that neither the specific choice of silicate 
optical constants nor the precise constituents of the ice mixtures will 
affect our conclusion since in the present paper, we are not intending to 
reproduce any specific observations but rather to demonstrate the effects 
of grain shape and the presence of an ice mantle and therefore to explore the 
validity of the spectrum subtraction method.  
We defer our detailed modelling of astronomical silicates 
to Zhao et al.\ (2001) where the chemical variations of 
silicate, carbonaceous, and ice materials,
and grain shapes and morphology (porosities) 
will be discussed in detail. 

Since there are no existing optical constants available for such an 
${\rm H_2O: CH_3OH: CO_2 = 1.3: 1: 1}$ mixture, we construct a set of 
synthetic refractive indices from the existing data for pure H$_2$O at 80K,
pure CH$_3$OH at 75K, pure CO$_2$ at 70K (Hudgins et al.\ 1993;
with mass densities $\approx 1.00, 1.23, 1.56\g\cm^{-3}$, respectively) 
by applying the Bruggeman Effective Medium Theory (EMT) 
(Bohren \& Huffman 1983) and assuming a volume fraction 
of ${\rm H_2O: CH_3OH: CO_2 = 1: 1.1: 1.2}$.
Theoretical spectra calculated from the synthetic optical constants are
consistent with the laboratory IR spectrum of Ehrenfreund et al.\ (1999).
Again, we emphasize here that the precise knowledge of the optical 
constants of ices is not critical since the purpose of this paper is 
not to infer the exact dust composition. 

\section{Spherical grains}
We first consider spherical grains. Under the condition of 
$x=2\pi a/\lambda\ll 1$ and $|mx|\ll 1$ ($a$ is the grain radius, 
$m$ is the grain refractive index which relates to the dielectric 
function $\epsilon$ through $\epsilon = m^2$), the absorption cross 
section of a core-mantle spherical grain $C_{\rm abs}$ can be 
calculated from the Rayleigh approximation (van de Hulst 1957)
\begin{equation}
C_{\rm abs}/V = 3 \frac{2\pi}{\lambda} {\rm Im}\left\{
             \frac{(\epsilon_{\rm m}-1)(\epsilon_{\rm c}+2\epsilon_{\rm m})+
             f_{\rm c}(2\epsilon_{\rm m}+1)(\epsilon_{\rm c}-\epsilon_{\rm m})}
             {(\epsilon_{\rm m}+2)(\epsilon_{\rm c}+2\epsilon_{\rm m})+
             f_{\rm c}(2\epsilon_{\rm m}-2)(\epsilon_{\rm c}-\epsilon_{\rm m})}
             \right\}
\end{equation}
where $V$ is the dust volume; $\epsilon_{\rm c}$ and $\epsilon_{\rm m}$ are
the complex dielectric functions for the core and mantle materials, 
respectively; $f_{\rm c}$ is the volume fraction of the core. The scattering
cross sections $C_{\rm sca}$ are much smaller than $C_{\rm abs}$ so that
the extinction cross sections $C_{\rm ext} \approx C_{\rm abs}$. 

We carried out calculations for spherical silicate core-ice mantle 
grains, taking the silicate core size $a_{\rm sil} = 0.07\mu$m, 
a typical size for the interstellar silicate dust (Li \& Greenberg 1997).
Note the knowledge of the exact grain size is not critical since 
interstellar grains ($\sim 0.1\mu$m) are in the Rayleigh limit 
in the wavelength range of interest here 
and also because of this no size distribution is needed. 

In Fig.\,1a we plot the absorption cross section of the core-mantle 
grain $C_{\rm abs}^{\rm sil+ice}$ with $a_{\rm sil}$=0.07$\mu$m, 
$f_{\rm sil}$=2/3 (the volume fraction of the silicate core,
$\equiv 1-V_{\rm ice}/V_{\rm sil+ice}$) in the silicate features region.
Also plotted are the absorption cross sections for a pure 
silicate grain $C_{\rm abs}^{\rm sil}$ with the same size as the silicate 
core, and from a pure ice grain $C_{\rm abs}^{\rm ice}$ with size 
$a_{\rm ice}\approx$0.0556$\mu$m corresponding to the volume of the ice 
mantle. Note that the pure ice grain spectrum closely reproduces the ice 
features of the core-mantle grain. According to the spectrum subtraction 
method, the contribution of the silicate core to the total absorption 
cross section 
$\approx C_{\rm abs}^{\rm sil-SSM} \equiv 
C_{\rm abs}^{\rm sil+ice}-C_{\rm abs}^{\rm ice}$. 
However, as clearly illustrated in Fig.\,1a, 
the $C_{\rm abs}^{\rm sil-SSM}$
spectrum deviates significantly from that of 
the pure silicate grain $C_{\rm abs}^{\rm sil}$: 
the peak position of 
both the 10$\mu$m Si-O and the 18$\mu$m O-Si-O features shift to 
longer wavelengths; both features have a broader red wing
with an error $\chi_{\rm SSM} \equiv 
C_{\rm abs}^{\rm sil-SSM}/C_{\rm abs}^{\rm sil}-1$
$\approx 8\%$ for the 10$\mu$m wing
and $\approx 12\%$ for the 18$\mu$m wing.
These differences are comparable to some of the differences ascribed by
Demyk et al.\ (1999) to different materials and/or different porosities
(for example, the simple spherical enstatite and olivine
fits to the silicate absorption spectra [{\it with ice absorptions
subtracted}] of RAFGL 7009S and IRAS 19110 which were rejected by 
Demyk et al.\ fail to reproduce the red wing by 
$< 10\%$ for $\lambda < 18\mu$m
and $< 20\%$ for $18< \lambda < 20\mu$m).
In light of this, Demyk et al.\ (1999)'s invoking of {\it an admixture of
iron, aluminum and some degree of porosity} to explain the red wing of the 
18$\mu$m feature seems unnecessary since it may be just an artifact of
the spectrum subtraction method they adopted (see their \S4.2 and Fig.\,4).

We have also considered the variation of the ice mantle thickness. 
For illustration we present in Fig.\,1b the errors $\chi_{\rm SSM}$ 
caused by the spectrum subtraction method 
for $f_{\rm ice}\equiv V_{\rm ice}/V_{\rm sil+ice}$ =1/5 (dotted), 
1/3 (solid), and 1/2 (dashed), respectively. 
Fig.\,1b shows that the spectrum subtraction method can not 
give acceptable results even for a very thin layer of 
ice mantle ($\approx$0.0054$\mu$m for
$f_{\rm ice}$ =1/5).

\begin{figure}
\psfig{figure=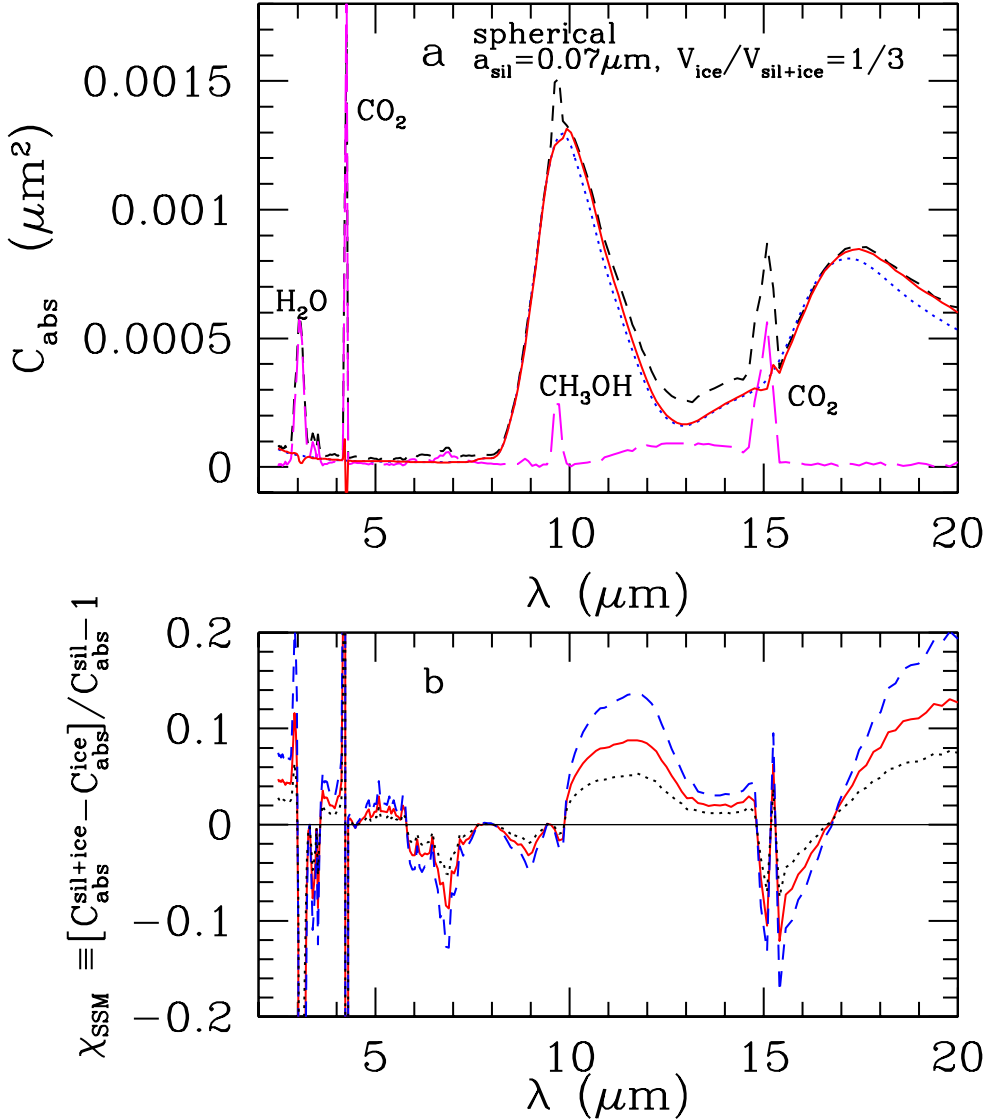,width=\figwidth}
\caption{
        {\bf a}: Absorption cross sections of spherical grains.
	Dashed ($C_{\rm abs}^{\rm sil+ice}$): silicate core-ice mantle 
        grains with a silicate core radius 0.07$\mu$m and a silicate
        volume fraction $f_{\rm sil} = 2/3$ (i.e., $f_{\rm ice}=1/3$);
        long dashed ($C_{\rm abs}^{\rm ice}$): the volume-equivalent sphere 
        for the ice mantle;
        dotted ($C_{\rm abs}^{\rm sil}$): a pure silicate sphere with the 
        same size as the silicate core of the core-mantle dust;
        solid: $C_{\rm abs}^{\rm sil-SSM}$.
        In the framework of the SSM, one should have 
        $C_{\rm abs}^{\rm sil-SSM} \approx C_{\rm abs}^{\rm sil}$.
        The contrast between $C_{\rm abs}^{\rm sil-SSM}$ (solid)
        with $C_{\rm abs}^{\rm sil}$ (dotted) indicates the 
        unapplicability of the SSM to spherical grains.
        {\bf b}: The errors $\chi_{\rm SSM}$ 
        caused by the spectrum subtraction method 
        for $f_{\rm ice}$ =1/5 (dotted), 
        1/3 (solid), and 1/2 (dashed).
        The errors in the red wings of the 10$\mu$m and 18$\mu$m features
        are comparable to the differences ascribed by
        Demyk et al.\ (1999) to different materials and/or 
        different porosities (see our text and their \S4.2 and Fig.\,4).
        The sharp features at 3.1, 4.2, 9.75, and 15.2$\mu$m
        are due to the incompleteness in subtracting the ice features.
        This will not affect our conclusion since in this work we are
        mainly interested in the broad 10 and 18$\mu$m silicate bands.
        }
\end{figure}

\section{Spheroidal grains}
Cosmic dust grains are unlikely to be spherical. On the contrary, 
polarization of starlight implies both nonsphericity and alignment. 
We now consider spheroidal shape which can be exactly solved in
the Rayleigh limit. For an ensemble of randomly oriented spheroidal 
grains\footnote{%
  Similar results are expected for aligned grains; 
  for example, the deviation 
  $\chi^{\prime} \equiv \sqrt{\sum\left\{\left[C_{\rm abs}^{\rm sil+ice} - 
  C_{\rm abs}^{\rm ice}\right]/C_{\rm abs}^{\rm sil}-1\right\}^2/N}$ 
  for the wavelength range of $8.5 \le \lambda \le 20 \mu$m 
  ($N$ is the number of data points)
  from 3:1 prolate grains of perfect spinning alignment 
  with $f_{\rm ice}$=1/3 is about 0.023
  in comparison with $\chi^{\prime}=0.020$ for randomly oriented prolates.
  },
the absorption cross section is (Greenberg 1972; Lee \& Draine 1985)
\begin{equation}
C_{\rm abs} = \frac{1}{3} C_{\rm abs}^{\|} + \frac{2}{3} C_{\rm abs}^{\bot}
\end{equation}
where $C_{\rm abs}^{\|}$ and $C_{\rm abs}^{\bot}$ are the absorption cross 
sections for light polarized parallel and perpendicular, respectively, to 
the grain symmetry axis. In the Rayleigh limit, for {\it confocal} 
core-mantle spheroidal grains we can calculate the absorption cross sections
$C_{\rm abs}^{\|, \bot}$ from (Gilra 1972; Draine \& Lee 1984)
\begin{equation}
C_{\rm abs}^{\|, \bot}/V = \frac{2\pi}{\lambda} {\rm Im}\left\{ 
             \frac{(\epsilon_{\rm m}-1)\left[L_{\rm c}^{\|, \bot}
             (\epsilon_{\rm c}-\epsilon_{\rm m})+\epsilon_{\rm m}\right]+
              f_{\rm c}(\epsilon_{\rm c}-\epsilon_{\rm m})
              \left[L_{\rm m}^{\|, \bot}(1-\epsilon_{\rm m})+
              \epsilon_{\rm m}\right]} {\left[L_{\rm c}^{\|, \bot}
              (\epsilon_{\rm c}-\epsilon_{\rm m})+\epsilon_{\rm m}\right]
              \left[1+L_{\rm m}^{\|, \bot}(\epsilon_{\rm m}-1)\right]+
              f_{\rm c}L_{\rm m}^{\|, \bot}(1-L_{\rm m}^{\|, \bot})
              (\epsilon_{\rm c}-\epsilon_{\rm m})(\epsilon_{\rm m}-1)}
              \right\}
\end{equation}
where $L_{\rm c}^{\|}$, $L_{\rm c}^{\bot}$, $L_{\rm m}^{\|}$, and
$L_{\rm m}^{\bot}$ are the depolarization factors of the core and mantle
parallel and perpendicular, respectively, to the grain symmetry axis.
The depolarization factors $L^{\|}$, $L^{\bot}$ relate to each other through 
$L^{\bot} = (1-L^{\|})/2$ because of rotational symmetry and characterize 
grain shape via Eq.(4): 
\begin{equation}
L^{\|} = \cases{\frac{1-e^2}{e^2} 
      \left[\frac{1}{2e} {\rm ln}\left(\frac{1+e}
      {1-e}\right)-1\right], &for prolate ($a>b$) \cr
      \frac{1+e^2}{e^2}\left(1-\frac{1}{e}
      {\rm tan^{-1}}e\right), &for oblate ($a<b$)\cr}
\end{equation}
where the eccentricity $e$ is given by the two semiaxes $a$ and $b$
(with $a$ semiaxis along the symmetry axis and $b$ semiaxis perpendicular
to the symmetry axis) 
\begin{equation}
e = \cases{\sqrt{1-(b/a)^2}, & for prolate ($a>b$) \cr
           \sqrt{(b/a)^2-1}, & for oblate ($a<b$) \cr}
\end{equation}
and the eccentricities of core and mantle $e_{\rm c}$, $e_{\rm m}$
($e_{\rm c} > e_{\rm m}$) relate to each other through
\begin{equation}
f_{\rm c} = \cases{\frac{1-e_{\rm c}^2}{e_{\rm c}^3} \frac{e_{\rm m}^3}
            {1-e_{\rm m}^2}, & for prolate ($a_{\rm m} > b_{\rm m}$) \cr
            \frac{1+e_{\rm c}^2}{e_{\rm c}^3} \frac{e_{\rm m}^3}
            {1+e_{\rm m}^2}, & for oblate ($a_{\rm m} < b_{\rm m}$) \cr}
\end{equation}
In the special case of $\epsilon_{\rm c}=\epsilon_{\rm m}=\epsilon$, 
Eq.(3) reduces to the solution for a homogeneous spheroidal grain 
\begin{equation}
C_{\rm abs}^{\|, \bot}/V = \frac{2\pi}{\lambda} {\rm Im}\left\{
            \frac{\epsilon-1}{(\epsilon-1)L^{\|,\bot}+1}\right\}
\end{equation}

By applying the method described above, we calculate the absorption cross 
sections of {\it confocal} silicate core-ice mantle spheroidal grains with 
a volume-equivalent sphere radius (the radius of the equivalent sphere having
the same volume as the spheroid) 
$r_{\rm sil}^{\rm eq} = (ab^2)^{1/3}$ = 0.07$\mu$m for the silicate core
and a variable ice mantle controlled by $f_{\rm sil}$.
We first consider $a/b=3$ prolate core-mantle grains 
with $r_{\rm sil}^{\rm eq}$ = 0.07$\mu$m ($a/b=3$ is for the ice mantle; 
the corresponding $a/b$ for the silicate core is 
$\approx 3.57$ if $f_{\rm sil}$=2/3). 
The reason for choosing an $a/b=3$ prolate shape is that 
Greenberg \& Li (1996) found that it provides the best fit to the
10, 18$\mu$m silicate polarization features of the Becklin-Neugebauer 
(BN) object in terms of ice-coated core-mantle dust grains.
Also calculated are 1) $C_{\rm abs}^{\rm ice}$, the absorption cross 
section of the volume-equivalent pure ice prolate ($a/b=3$); 
2) $C_{\rm abs}^{\rm sil}$, the absorption cross section of 
the volume-equivalent pure silicate prolate 
with $a/b\approx 3.57$ determined from Eqs.(5,6); 
and 3) the ice mantle ``subtracted'' silicate absorption 
$C_{\rm abs}^{\rm sil-SSM}$. 
In Fig.\,2a we plot the corresponding errors $\chi_{\rm SSM}$
for $f_{\rm ice}$ =1/5 (dotted), 1/3 (solid), and 1/2 (dashed).

As can be seen in Fig.\,2a, $C_{\rm abs}^{\rm sil-SSM}$ is closer to 
$C_{\rm abs}^{\rm sil}$ in comparison with the spherical model (\S3). 
The errors $\chi_{\rm SSM}$ for the red wings of the 10$\mu$m 
and 18$\mu$m features are respectively $\sim 2.5\%$ and 
$\sim 2\%$ (for $f_{\rm sil}$=2/3),
in sharp contrast to those of the spherical model
($\sim 8\%$ and $\sim 12\%$; see Fig.\,1b).  

Similarly, we carried out the same calculations but for 
oblate grains with $a/b=1/2$ (the core has $a/b\approx 1/2.52$). 
The $a/b=1/2$ oblate shape was chosen since it was shown to provide 
a better match than any other shapes to the 3.1$\mu$m ice polarization 
feature (Lee \& Draine 1985) and the 10$\mu$m silicate polarization 
feature (Hildebrand \& Dragovan 1995) in terms of ice-coated silicate 
grains (without an intermediate carbonaceous mantle). 
In Fig.\,2b we plot $\chi_{\rm SSM}$  
for $f_{\rm ice}$ =1/5 (dotted), 1/3 (solid), 
and 1/2 (dashed) for oblate grains. 
The results are very similar to the $a/b=3$ prolate model (Fig.\,2a).
The errors are $< 3\%$ for the red wings of 
the 10$\mu$m and 18$\mu$m features.  

From these results we conclude that for spheroidal grains, 
the spectrum subtraction method works better than for spherical 
grains\footnote{%
  Why the spheroid model works better is not obvious 
  but it may be related to the fact that the spheroidal absorptivity 
  follows the ``material'' absorptivity ($m^{''}$) while spherical 
  one does not (Greenberg et al.\ 1983; van de Bult et al.\ 1985).
  }.
Lacking prior knowledge of the {\it true} shape of dust grains,
the spectrum subtraction method, as a first approximation,
can be adopted to study the silicate mineralogy if grains 
are taken to be spheroidal.
This does not necessarily imply that grains need to be
spheroidal. What it does imply is that the errors caused by 
the spectrum subtraction method, in reducing the silicate
mineralogy in terms of a spherical shape, are larger than
those of spheroidal shapes. 
If the choice of spheroidal shapes seems rather arbitrary
despite the fact that the error $\chi_{\rm SSM}$ is indeed smaller, 
we remind the reader that the spectrum subtraction method
is physically incorrect. A better approach to 
the silicate mineralogy problem is to carry out 
iterative modelling of the observational spectra
using Mie theory together with the effective medium theory
or the discrete-dipole approximation for various shapes
(e.g. see Zhao et al.\ 2001).

\begin{figure}
\psfig{figure=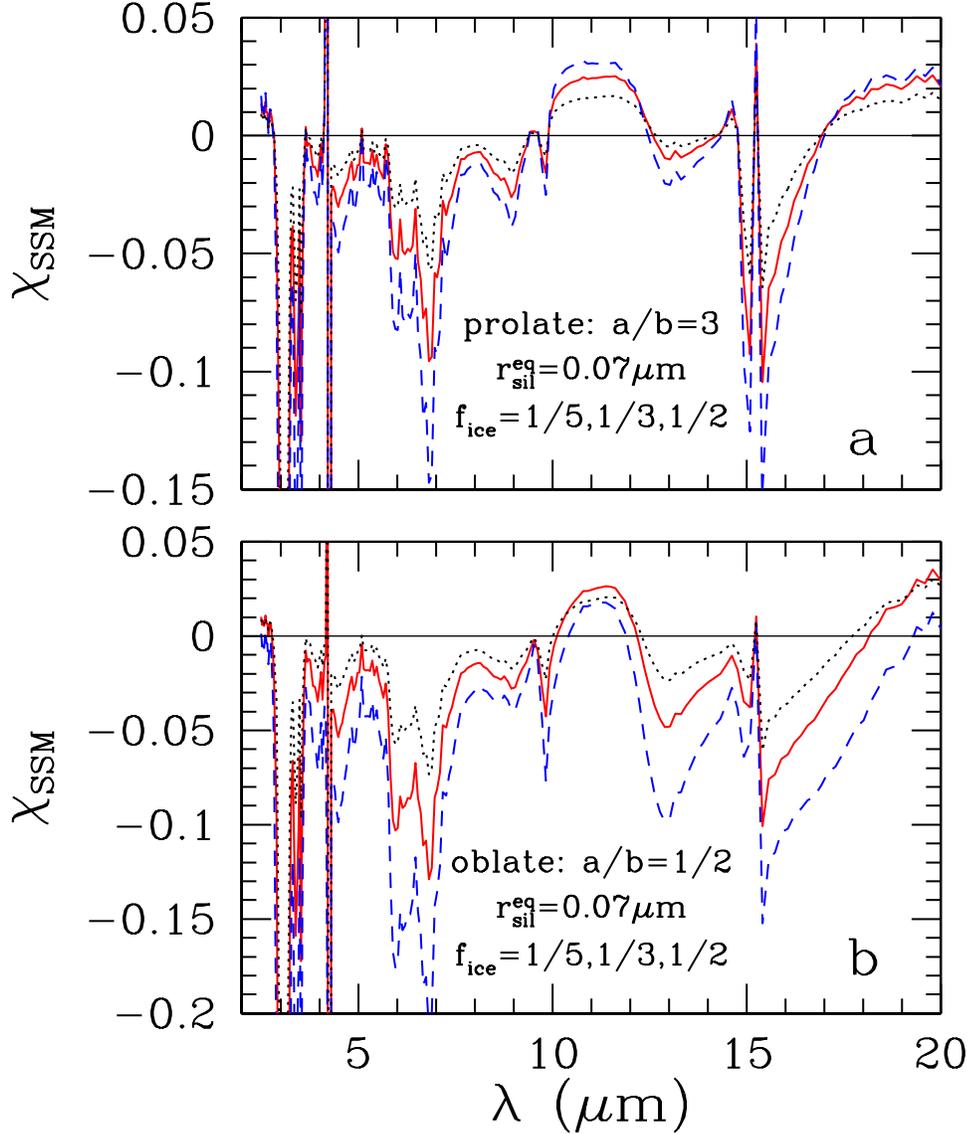,width=\figwidth}
\caption{
        \label{fig:prolate}
        The errors $\chi_{\rm SSM}$ 
        caused by the spectrum subtraction method 
        for $f_{\rm ice}$=1/5 (dotted), 1/3 (solid), and 1/2 (dashed)
        -- {\bf a}: for $a/b=3$ prolate;
        {\bf b}: for $a/b=1/2$ oblate.
        In comparison with the spherical basis, the spectrum subtraction 
        method based on spheroidal grains provides a better approach.
        The errors at the red wings of the 10$\mu$m and 18$\mu$m features
        are much smaller than those of the spherical model (see Fig.\,1b).
        }
\end{figure}

\section{Grains with a continuous distribution of ellipsoids (CDE)}
The shape of a spheroidal grain is characterized by its eccentricity $e$
(and the depolarization factors $L^{\parallel}$, $L^{\bot}$ via Eq.[4]). 
We now extend \S4 where only a single shape (a single eccentricity) is 
considered to an average of a distribution of shapes (for the same volume 
grain). We assume two kinds of shape distribution functions: 
1) $dP/dL^{\parallel}$ = constant, i.e., all shapes are equally probable
(Bohren \& Huffman 1983); 
2) $dP/dL^{\parallel} = 12 L^{\parallel}[1-L^{\parallel}]^2$  
(Ossenkopf, Henning, \& Mathis 1992; Will \& Aannestad 1999) 
which peaks at spheres ($L^{\|}=L^{\bot}$=1/3) and is symmetric 
about spheres with respect to eccentricity and drops to zero for 
the extreme cases ($e\rightarrow 1$ [$L^{\parallel}\rightarrow 0$]
infinitely thin needles or $e\rightarrow \infty$ 
[$L^{\parallel}\rightarrow 1$] infinitely flattened pancakes). 
Averaging over the shape distribution, 
we have the resultant absorption cross section 
$C_{\rm abs} = \int_{0}^{1} dL^{\parallel} dP/dL^{\parallel} 
C_{\rm abs}(L^{\parallel})$ where $C_{\rm abs}(L^{\parallel})$ is 
the absorption cross section of a particular shape $L^{\parallel}$ 
(note $L^{\parallel}$ is for the mantle;
the core depolarization factor is derived from Eqs.[4-6]).
Again, we assume confocal geometry for core-mantle grains, 
with the above $dP/dL^{\parallel}$ (as well as $e$) applying 
to the outer surface.

We calculated the averaged absorption cross sections 
of grains with a continuous distribution of ellipsoids (CDE) with 
$r_{\rm sil}^{\rm eq}$=0.07$\mu$m, $f_{\rm sil}$=2/3
for $dP/dL^{\parallel}$ = constant and 
$dP/dL^{\parallel} = 12 L^{\parallel}[1-L^{\parallel}]^2$ respectively. 
We also calculated (1) $C_{\rm abs}^{\rm sil}$, the averaged absorption 
cross sections predicted from the silicate core itself with  
eccentricities obtained from Eq.(6) for a given set of mantle 
eccentricities $e$ and $f_{\rm sil}$; (2) $C_{\rm abs}^{\rm ice}$, 
the averaged absorption cross sections of ice spheroids with the same 
volumes and the same eccentricities $e$ as the ice mantles of 
the core-mantle spheroids; and (3) $C_{\rm abs}^{\rm sil-SSM}$.

It is found that the $C_{\rm abs}^{\rm sil-SSM}$ of the 
$dP/dL^{\parallel}$ = constant model gives similar results to those of 
the single-shaped spheroidal grains (\S4) and leads to smaller 
errors than the spherical model (see Fig.\,3a). 
In contrast, although the 
$dP/dL^{\parallel} = 12 L^{\parallel}[1-L^{\parallel}]^2$ model
was generally considered as ``more physically reasonable'' 
(e.g. see Ossenkopf et al.\ 1992),
it results in a much larger error 
(even for $f_{\rm ice} = 1/5$, see Fig.\,3b). 
In particular, $C_{\rm abs}^{\rm sil-SSM}$ deviates from 
$C_{\rm abs}^{\rm sil}$ significantly in the peak regions
of the 10, 18$\mu$m. This can be understood by the fact that 
the $dP/dL^{\parallel} = 12 L^{\parallel}[1-L^{\parallel}]^2$ 
distribution peaks at {\it spheres} for which the spectrum subtraction 
method results in a significant error (\S3). In comparison with \S4, 
we conclude that the CDE model does not seem to be better 
suitable to be applied within the framework of 
the spectrum subtraction method than the single 
(non-spherical) shape model. 

\begin{figure}
\psfig{figure=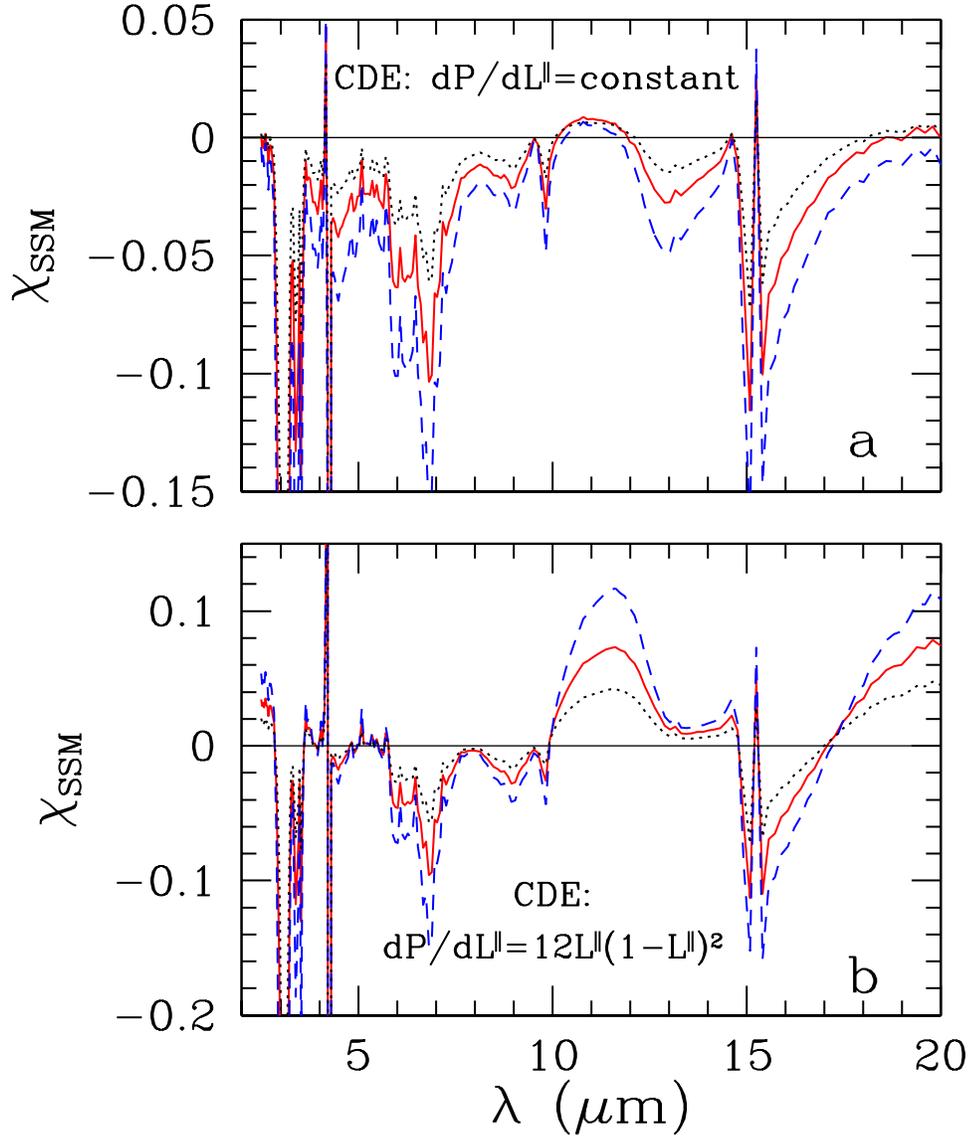,width=\figwidth}
\caption{
        \label{fig:err}
        The errors $\chi_{\rm SSM}$ 
        caused by the spectrum subtraction method 
        for $f_{\rm ice}$=1/5 (dotted), 1/3 (solid), and 1/2 (dashed)
        -- {\bf a}: for a continuous distribution 
        of spheroids with all shapes equally probable 
        (i.e. $dP/dL^{\parallel}$=constant);
        {\bf b}: for a continuous distribution of spheroids
        with the shape probability distribution 
        $dP/dL^{\parallel} = 12 L^{\parallel}[1-L^{\parallel}]^2$.
	The silicate core size is always taken to be
        $r_{\rm sil}^{\rm eq}$=0.07$\mu$m.
        }
\end{figure}

\section{Coated core-mantle grains}
Finally, we consider a model in which the silicate core is coated by an 
carbonaceous mantle (in addition to its outer ice mantle) (Greenberg 1978). 
We emphasize here the fact that with or without ice mantles
there is generally no pure silicate core but rather one which bears an
organic mantle as seen in the diffuse interstellar medium, cometary dust,
and as implied in molecular clouds (see Greenberg \& Li 1999 for a summary).
The absorption cross sections for an {\it equal-eccentricity} three-layered
silicate core-carbonaceous inner mantle-ice outer mantle grain is 
(Farafonov 2001; Voshchinnikov \& Mathis 1999)
\begin{equation}
C_{\rm abs}^{\|, \bot}/V = \frac{2\pi}{\lambda} \frac{4\pi}{3}
         {\rm Im}\left\{\frac{A_2^{\|,\bot}-A_1^{\|,\bot}} 
         {(A_2^{\|,\bot}-A_1^{\|,\bot}) L^{\|,\bot}+A_1^{\|,\bot}}\right\}
\end{equation}

\begin{equation}
\begin{array}{rcl}
\left(\begin{array}{c} A_1^{\|,\bot}\\A_2^{\|,\bot} \end{array}\right) & = &
\left(\begin{array}{cc}1 & L^{\|,\bot}\\ 
\epsilon_{\rm om} & \epsilon_{\rm om}(L^{\|,\bot}-1)\end{array}\right) 
\left(\begin{array}{cc}(\epsilon_{\rm im}/\epsilon_{\rm om}-1)L^{\|,\bot}+1 & 
(\epsilon_{\rm im}/\epsilon_{\rm om}-1)L^{\|,\bot}(L^{\|,\bot}-1)/f_{\rm im}\\ 
-(\epsilon_{\rm im}/\epsilon_{\rm om}-1)f_{\rm im} & 
-(\epsilon_{\rm im}/\epsilon_{\rm om}-1)
(L^{\|,\bot}-1)+1\end{array}\right) \\
~ & ~ & \left(\begin{array}{c} 
        (\epsilon_{\rm c}/\epsilon_{\rm im}-1)L^{\|,\bot}+1\\
-(\epsilon_{\rm c}/\epsilon_{\rm im}-1)f_{\rm c}\end{array}\right) 
\end{array}
\end{equation}
where $\epsilon_{\rm c}$, $\epsilon_{\rm im}$, $\epsilon_{\rm om}$ are the
dielectric functions for the silicate core, the organic refractory inner
mantle, the ice outer mantle, respectively; $f_{\rm c}$ and $f_{\rm im}$
are the volume fractions of the silicate core and the organic refractory
inner mantle, respectively. Note that, in contrast to \S4, the depolarization
factors here are the same for all layers.

We have calculated the absorption cross sections of
ice-coated silicate core-organic refractory mantle grains with 
$r_{\rm sil}^{\rm eq}$=0.07$\mu$m, $f_{\rm sil}=f_{\rm or}$=2/5,
$f_{\rm ice}$=1/5, for $a/b=3$ prolate as well as for $a/b=1/2$ oblate,
where $f_{\rm sil}$ ($\equiv f_{\rm c}$), 
$f_{\rm or}$ ($\equiv f_{\rm im}$),
and $f_{\rm ice}$ ($\equiv 1-f_{\rm c}-f_{\rm im}$)
are respectively the fractional volumes of the silicate core,
the inner organic refractory mantle, and the outer ice mantle. 
As not unexpected, $C_{\rm abs}^{\rm sil-SSM}$
($\equiv C_{\rm abs}^{\rm sil+or+ice}-C_{\rm abs}^{\rm ice}$) 
is very different from $C_{\rm abs}^{\rm sil}$ in the peak positions, 
the feature widths, and the feature strengths of both the 10$\mu$m and 
18$\mu$m features. Since the organic refractory material has little 
absorption around that region, the inclusion of such an inner mantle 
significantly suppresses the 10$\mu$m feature (not shown here). 
It is obvious that silicate grains with carbonaceous coating 
will not behave like naked silicates.
Therefore, it is essential to take into account 
the possible heterogeneous nature of cosmic dust
when modelling the silicate mineralogy.

\section{Summary}
We have investigated the validity of the ``{\it spectrum subtraction}''
method or the ``{\it absorption additivity assumption}'' widely adopted 
in studies of astronomical silicate mineralogy when interpreted using 
silicate core-ice mantle grains with spherical, spheroidal, 
and a continuous distribution of ellipsoids (CDE) shapes.  
We found that: 
1) the spectrum subtraction method results in significant errors for 
spherical grains; 
2) for spheroidal grains, the errors are much less significant and
the spectrum subtraction method is reasonably successful in removing 
the ice mantle effects;
3) grains with a uniform distribution of spheroidal shapes 
produce results little different from those for a single eccentricity;
4) as expected, grains with sphere-peaked distribution of spheroidal 
shapes, lead to appreciable errors.
In addition, we have also discussed the effects caused by the inclusion 
of an intermediate carbonaceous mantle which is often not taken into 
account in silicate considerations. As not unexpected, the spectrum 
subtraction method applied here leads to even more unacceptable errors.

Our conclusion is that, in order to correctly infer the precise 
composition of astronomical silicates, care should be taken to 
appropriately take into account the heterogeneous (chemical and/or 
structural) nature of dust grains. The spectrum subtraction method
should be used together with the non-spherical (e.g., prolate or oblate)
assumption. It does not appear to have obvious advantages to use 
a distribution of spheroidal shapes.

\section*{Acknowledgments}
A. Li and G. Zhao were deeply saddened by the passing away of 
Prof. J. Mayo Greenberg (the second author of this paper) 
on November 29, 2001. As a pioneer in the fields of 
cosmic dust, comets, astrochemistry, astrobiology 
and light scattering, Mayo will be remembered forever. 
It was a great experience for A. Li to work with Mayo in Leiden.
We thank Prof. B.T. Draine for his invaluable comments and suggestions;
Prof. J.I. Lunine for helpful discussions;
Prof. J.S. Mathis for clarification;
and Dr. R.H. Lupton for the availability of the SM plotting package. 
We are grateful to the anonymous referee for helpful advice. 
We are also grateful to Drs. K. Demyk and L. d'Hendecourt
for providing us with their RAFGL 7009S and IRAS 19110 spectra.
This research was supported in part by NASA grant NAG5-7030 and 
NSF grant AST-9619429, and the World Laboratory E-17 Project 
``Dust and Gas Chemistry in Star-forming Regions''.


\begin{thebibliography}{}
\bibitem[]{}Bohren, C.F., \& Huffman, D.R. 1983, Absorption and Scattering 
            of Light by Small Particles (New York: Wiley)
\bibitem[]{}Bradley, J.P., Keller, L.P., Snow, T.P., et al.\
            1999, Science, 285, 1716 
\bibitem[]{}Brucato, J.R., Colangeli, L., Mennella, V., Palumbo, P.,
            \& Bussoletti, E. 1999, Planet. Space Sci., 47, 773
\bibitem[]{}Dartois, E., Demyk, K., d'Hendecourt, L., \& Ehrenfreund, P. 
            1999, A\&A, 351, 1066
\bibitem[]{}Demyk, K., Jones, A.P., Dartois, E., Cox, P., 
            \& d'Hendecourt, L. 1999, A\&A, 349, 267
\bibitem[]{}Dorschner, J. 1999, in Formation and Evolution of Solids in 
            Space, ed. J.M. Greenberg \& A. Li (Dordrecht: Kluwer), 229
\bibitem[]{}Draine, B.T., \& Lee, H.M. 1984, ApJ, 285, 89
\bibitem[]{}Ehrenfreund, P., Kerkhof, O., Schutte, W.A., et al.\
            1999, A\&A, 350, 240
\bibitem[]{}Farafonov, V.G. 2001, Opt. Spectrosc., 90, 574
\bibitem[]{}Gilra, D.P. 1972, in The Scientific Results from the OAO-2,
            ed. A.D. Code (NASA SP-310), 295 
\bibitem[]{}Greenberg, J.M. 1972, J. Colloid Interface Sci., 39, 513
\bibitem[]{}Greenberg, J.M. 1978, in Cosmic Dust, ed. J.A.M. McDonnell
            (New York: Wiley), 187 
\bibitem[]{}Greenberg, J.M. 1982, in Comets, ed. L.L. Wilkening 
            (Tucson: Univ. Arizona Press), 131
\bibitem[]{}Greenberg, J.M., van de Bult, C.E.P.M., \& Allamandola, L.J.
            1983, J. Phys. Chem., 87, 4243
\bibitem[]{}Greenberg, J.M., \& Li, A. 1996, A\&A, 309, 258
\bibitem[]{}Greenberg, J.M., \& Li, A. 1999, Adv. Space Res., 24, 497
\bibitem[]{}Hildebrand, R.H., \& Dragovan, M. 1995, ApJ, 450, 663
\bibitem[]{}Hudgins, D.M., Sandford, S.A., Allamandola, L.J., 
            \& Tielens, A.G.G.M. 1993, ApJS, 86, 713
\bibitem[]{}Kissel, J., Brownlee, D.E., B\"{u}chler, K., et al.\ 1986,  
            Nature, 321, 336
\bibitem[]{}Lee, H.M., \& Draine, B.T. 1985, ApJ, 290, 211
\bibitem[]{}Li, A., \& Greenberg, J.M. 1997, A\&A, 323, 566
\bibitem[]{}Malfait, K., Waelkens, C., Waters, L.B.F.M., et al.\
            1998, A\&A, 332, L25
\bibitem[]{}Mathis, J.S., \& Whiffen, G. 1989, ApJ, 341, 808
\bibitem[]{}Mathis, J.S. 1998, ApJ, 497, 824
\bibitem[]{}Ossenkopf, V., Henning, Th., \& Mathis, J.S. 1992, A\&A, 261, 567
\bibitem[]{}van de Bult, C.E.P.M., Greenberg, J.M., \& Whittet, D.C.B.
            1985, MNRAS, 214, 289
\bibitem[]{}van de Hulst, H.C. 1957, Light Scattering by Small Particles
            (New York: Wiley)            
\bibitem[]{}Voshchinnikov, N.V. \& Mathis, J.S. 1999, ApJ, 526, 257 
\bibitem[]{}Will, L.M., \& Aannestad, P.A. 1999, ApJ, 526, 242
\bibitem[]{}Wooden, D.H., Harker, D.E., Woodward, C.E., et al.\
            1999, ApJ, 517, 1034	
\bibitem[]{}Zhao, G., Li, A., \& Greenberg, J.M. 2001, in preparation  
\end{thebibliography}
\end{document}